\documentclass[]{aa}
\usepackage{graphicx}

\begin{document}

\title{The diffuse radio emission from the Coma cluster at 2.675~GHz and 4.85~GHz}
\titlerunning{The diffuse radio emission from the Coma cluster}

\author{M. Thierbach\inst{1,2}
\and
U. Klein\inst{3}
\and
R. Wielebinski\inst{1}
       }

\offprints{M.~Thierbach,
\email{thierb@mpifr-bonn.mpg.de}}

{\institute{Max--Planck--Institut f{\"u}r Radioastronomie, Auf dem H{\"u}gel 69,
53121 Bonn, Germany
\and
SMTO, Steward Observatory, The University of Arizona, 933 N.~Cherry Avenue, Tucson, Arizona 85721, USA
\and
Radioastronomisches Institut der Universit{\"a}t Bonn, Auf dem H{\"u}gel 71, 53121
Bonn, Germany
          }

\date{Received ?? / Accepted ??}

\abstract{
We present new measurements of the diffuse radio halo emission from the Coma cluster of galaxies at 2.675~GHz and 4.85~GHz using the Effelsberg 100--m telescope. After correction for the contribution from point sources we derive the integrated flux densities for the halo source (Coma~C), $S_{\rm 2.675~GHz}=(107\pm 28)~{\rm mJy}$ and $S_{\rm 4.85~GHz}=(26\pm 12)~{\rm mJy}$. These values verify the strong steepening of the radio spectrum of Coma~C at high frequencies. Its extent strongly depends on frequency, at 4.85~GHz it is only marginally visible. The measurement at 4.85~GHz is the first flux density determination for Coma~C at this high frequency. In order to quantify the spectral steepening we compare the expectations for the spectrum of Coma~C with the observations, resorting to basic models for radio halo formation. The in--situ acceleration model provides the best fit to the data. From equipartition assumptions we estimate a magnetic field strength $B_{\rm eq}=0.57\,(1+k)^{0.26}\,{\rm \mu G}$ in the intracluster medium of Coma, where k is the energy ratio of the positively and negatively charged particles.
\\
As a by--product of the 2.675~GHz observation we present a new flux density for the diffuse emission of the extended source 1253+275 ($S_{\rm 2.675GHz}=112\pm 10$). This measurement provides a smaller error range for the power--law fit to the spectrum ($\alpha=1.18\pm 0.02$) compared to previous investigations and yields an equipartition magnetic field strength of $B_{\rm eq}=0.56\,(1+k)^{0.24}\,{\rm \mu G}$.
 }
\maketitle
%
%________________________________________________________________

\section{Introduction}
The diffuse, extended radio emission of the Coma cluster of galaxies (Abell cluster number 1656) was detected by Large et al.~(\cite{large59}) and first investigated by Willson~(\cite{willson70}). The prominent diffuse, extended radio source Coma~C is the prototype of a cluster--wide radio halo. It was thoroughly investigated during the past years (e.g.~Schlickeiser et al.~\cite{schlickeiser87} (hereafter referred to as SST), Kim et al.~\cite{kim90}, Giovannini et al.~\cite{gio93}, Deiss et al.~\cite{deiss97}). In the frequency range below 1.4~GHz there are many observations available that suggest a consistent power--law shape of the radio spectrum. SST present measurements of the halo source using the 100--m Effelsberg telescope\footnote{The 100--m telescope at Effelsberg is operated by the Max--Planck--Institut f{\"u}r Radioastronomie (MPIfR) on behalf of the Max--Planck--Gesellschaft (MPG).}
%Based on observations with the 100--m telescope of the 
%MPIfR (Max--Planck--Institut f{\"u}r Radioastronomie) at Effelsberg.
 at 2.7~GHz. They found an integrated flux density significantly below that expected from the extrapolation from lower frequencies. The authors claim the existence of a spectral steepening of the radio spectrum of Coma~C at frequencies above 1~GHz. Comparing predictions of the spectral shape of the three basic models for the radio halo formation (primary, secondary electron model, in--situ acceleration model) SST favour the latter for the Coma cluster. Observations at 1.4~GHz by Deiss et al.~(\cite{deiss97}) with the Effelsberg telescope show the halo emission to be very extended. The subtraction of the contribution of point sources from the map results in a weak, diffuse radio component extended by more than 80$\arcmin$ in east--western direction, and by about 45$\arcmin$ north--south. The integrated flux density of this structure fits much better to the low--frequency part of the spectrum than to the 2.7~GHz point obtained by SST. Deiss et al.~deemed the  extremely sharp spectral break above 1.4~GHz as suggested by their own and the  SST data to be unrealistic. The authors assume that either the 2.7~GHz map suffers from incorrect baselines, or that the mapped area was too small to cover the whole halo source. 
The shape of the radio spectrum provides constraints on the physical background of the halo formation. The work of Deiss et al.~stimulated new sensitive observations of the diffuse emission in the Coma cluster at high frequencies ($>$1.4~GHz) in order to be able to derive constraints on the theory. The Effelsberg telescope is especially powerful to detect weak, extended emission.
 
\begin{figure*}[!t]
\resizebox{\hsize}{!}{\includegraphics{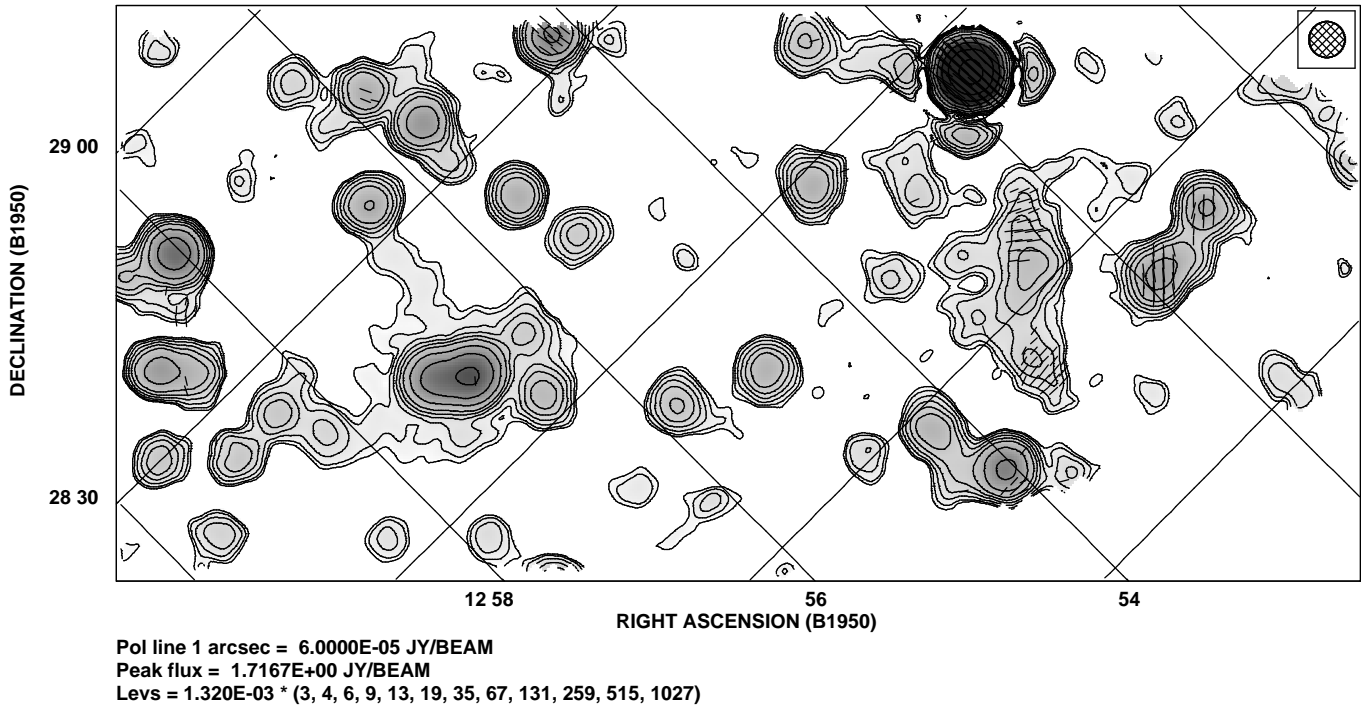}}
\caption{The Coma cluster of galaxies at 2.675~GHz. NE is to the left. The lowest contour marks the 3$\sigma$ level. The polarized intensity is represented by the length of the E--vectors. Note that the cluster center ($\alpha=12^{\rm h}57^{\rm m}, \delta=28\degr12'$) does not show any significant polarization.}
 
\label{fig11}
 
\end{figure*}
 
The Coma cluster of galaxies is located at a redshift $z=0.0232$ (Struble \& Rood \cite{struble91}), which implies a distance to the target of $\sim\!90\,h_{\rm 75}\!\!^{-1}$~Mpc and an angular scale of $\sim\!26\,h_{\rm 75}\!\!^{-1}$~kpc/arcmin (we adopt $H_0\!\!=\!\!75$~km/sec/Mpc throughout the paper).
%__________________________________________________________________
\section{Observations and data analysis}

The observations reported here were carried out with the Effelsberg 100--m radio telescope between July 1998 and May 1999. We mapped the Coma cluster  of galaxies at the frequencies of 2.675~GHz ($\lambda11.2$cm) and 4.85~GHz ($\lambda6.2$cm). Both receiver systems are installed in the secondary focus of the telescope. Their main properties can be found in Table~\ref{tabre}.

\begin{table}
\centering
\caption{Receiver parameters.}
\label{tabre}
\begin{tabular}{lcc}
\hline
& $2.675\,{\rm GHz}$ & $4.85\,{\rm GHz}$\\
\hline
Number of channels & $2$ & $4$\\
Number of horns & $1$ & $2$\\
System temperature $[{\rm K}]$ & $45$ & $30$\\
Bandwidth $[{\rm MHz}]$ & $40$ & $500$\\
HPBW $[\arcmin]$ & $4.3$ & $2.45$\\
HPBW $[\arcsec]$ & $258$ & $147$
\\
\hline
\end{tabular}
\end{table}

We scanned the telescope in azimuth at both frequencies. The three Stokes parameters $I$, $U$ and $Q$ were recorded simultaneously. We obtained several coverages of the cluster area. At 2.675~GHz we mapped a large area, covering the  central region and the extended source 1253+275. For the central region we added a couple of additional coverages. At 4.85~GHz we observed the center of the Coma cluster. The data reduction followed standard procedures for continuum observations with the 100--m telescope. For details of the observations consult Table~\ref{tabmaps}. Table~\ref{tabcal} lists the calibration sources together with the adopted parameters.

\setlength{\tabcolsep}{0.7mm}
\begin{table}[b]
\caption{Observational parameters. ($^{\rm a}$ Integration time per independent point)}
\label{tabmaps}
\begin{tabular}{ccccccc}
\hline
Freq. & \multicolumn{2}{c}{Center of map} & \multicolumn{1}{c}{Final} & \multicolumn{1}{c}{Integr.} & rms & see \\
& \multicolumn{1}{c}{$\alpha_{1950}$} & \multicolumn{1}{c}{$\beta_{1950}$} & \multicolumn{1}{c}{map size} & \multicolumn{1}{c}{time$^{a}$} & noise & Fig. \\
$[$GHz$]$ & $[\, ^h\quad ^m\quad ^s]$ & $[\,\degr \quad\arcmin \quad\arcsec]$ & \multicolumn{1}{c}{$[\,'\times \;']$} & $[$sec$]$ & $[$mJy/b.a.$]$ & \\
\hline
2.675 & 12 54 40 & 27 55 00 & $\!\!\!$150$\times$63 & 11 & 1.32 & \ref{fig11} \\
2.675 & 12 56 57 & 28 10 14 &  59$\times$58  & 33 & 1.10 & \ref{fig11c} \\
4.850 & 12 57 30 & 28 20 00 &  90$\times$90  &  8 & 0.67 & \ref{fig6}a \\
\hline
\end{tabular}
\end{table}

\begin{table}[b]
\caption{Properties of the calibration sources used. 3C295 and 3C309.1 were used only for total power calibration. $S$ stands for the flux density, $P$ for the degree of polarization, and $\psi$ marks the polarization angle.}
\label{tabcal}
 
\begin{center}
\begin{tabular}{l|ccc|ccc|c|c}
\hline
\multicolumn{1}{c|}{} & \multicolumn{3}{c|}{3C138} & \multicolumn{3}{c|}{3C286} & \multicolumn{1}{c|}{$\;$3C295$\;$} & \multicolumn{1}{c}{$\;$3C309.1} \\
& $S$ & $P$ & $\psi$ & $S$ & $P$ & $\psi$ & $S$ & $S$ \\
& $[{\rm Jy}]$ & $[\%]$ & $[\degr]$ & $[{\rm Jy}]$ & $[\%]$ & $[\degr]$ & $[\rm Jy]$ & $[{\rm Jy}]$ \\
\hline
2.675~GHz & $6.3$ & $10.1$ & $171$ & $11$ & $9.9$ & $33$ & $12.3$ & $4.8$\\
4.850~GHz & $3.8$ & $11.1$ & $169$ & $7.5$ & $11.3$ & $33$ & $6.6$ & $3.2$ \\
\hline
\end{tabular}
\end{center}
\end{table}
%________________________________________________________________

In Fig.~\ref{fig11} the radio map at 2.675~GHz showing the total intensity and the polarized emission is displayed. Fig.~\ref{fig11c} zooms the center of the cluster. This map results from 33 coverages of the area, yielding an rms noise of 1.1~mJy/beam. The cluster center does not show any significant linear polarization at 2.675~GHz.
 
\begin{figure}[!b]
\resizebox{\hsize}{!}{\includegraphics{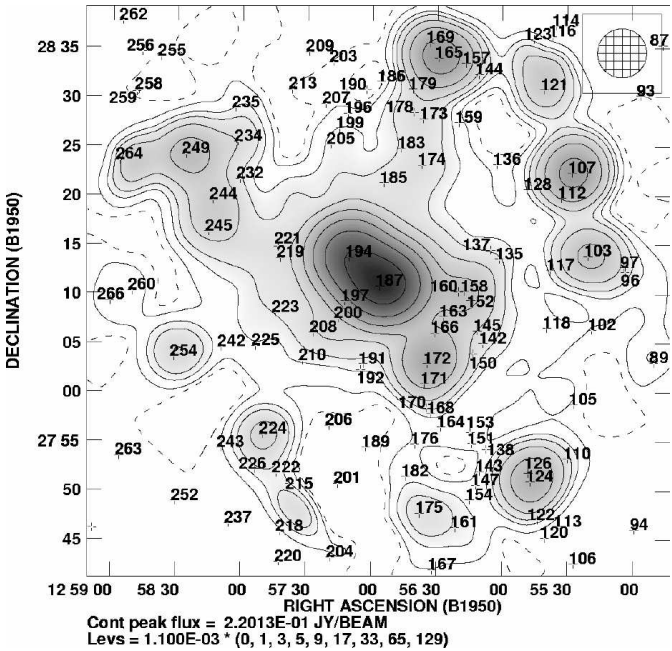}}
\caption{The central region of the Coma cluster of galaxies at 2.675~GHz smoothed to a resolution of $5'$ in order to improve the signal--to--noise of the diffuse emission. The dashed contour marks the zero level. The numbers mark positions of point--like sources, taken from the list of Kim~(\cite{kim94}).}
\label{fig11c}
\end{figure}
 
%________________________________________________________________

The map of the cluster at 4.85~GHz is shown in Fig.~\ref{fig6}. One can clearly see the polarized emission of the central galaxies NGC~4869 ($\alpha=12^{\rm h}56^{\rm m}56\fs5, \delta=28\degr10'50''$) and NGC~4874 ($\alpha=12^{\rm h}57^{\rm m}10\fs7, \delta=28\degr13'46''$). Note that no extended emission above the 3$\sigma$ level is seen. Nevertheless the map reveals slightly extended emission at a level of 0.67 to 1.34$\,$mJy/beam ($\stackrel{\wedge}{=} 1\ldots2\sigma$). In order to make it visible, the central part of the map is displayed in Fig.~\ref{fig6}{\bf b} down to zero flux. One can see extended structure around the two central galaxies.
 
\begin{figure}[!ht]
\resizebox{\hsize}{!}{\includegraphics{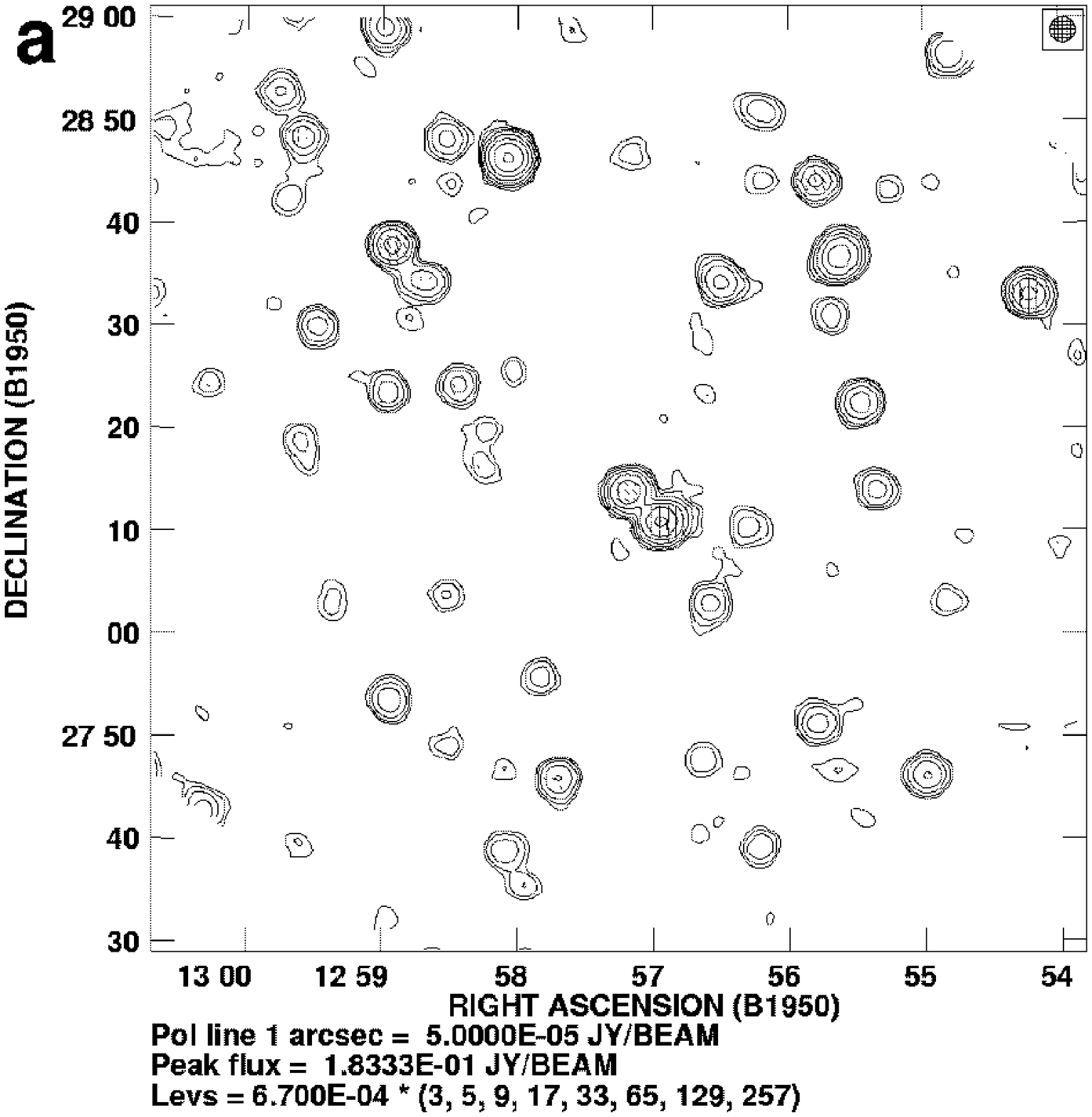}}

\vspace{5mm}

\resizebox{\hsize}{!}{\includegraphics{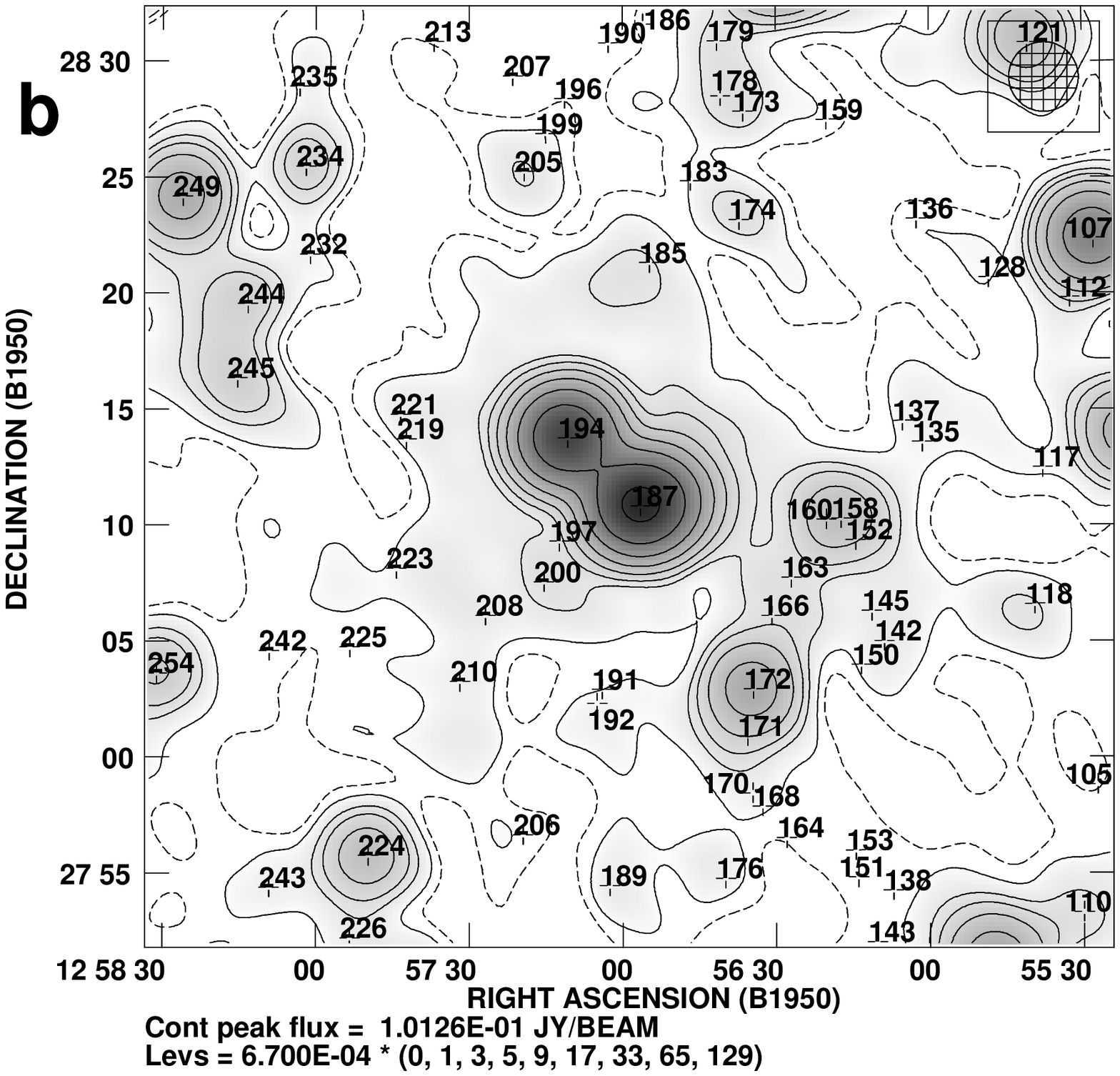}}
\caption{The Coma cluster of galaxies at 4.85~GHz. The top panel ({\bf a}) shows the complete map. The lowest contour marks the 3$\sigma$ level. The polarized intensity is represented by the length of the E--vectors. In the lower panel ({\bf b}) the central part of the map is shown. The data are smoothed to a resolution of $3'$ in order to improve the signal--to--noise of the diffuse emission. The dashed contour marks the zero level. The numbers mark the positions of the point--like sources from the list of Kim (\cite{kim94}).}
\label{fig6}
\end{figure}

%________________________________________________________________

\section{Results}
\subsection{Morphology}
\subsubsection*{The emission at 2.675~GHz}
The 2.675~GHz map shows various point sources, emission by galaxies, extended diffuse emission in the center of the cluster, and the extended diffuse the source 1253+275 (see Sect.~\ref{sect1253}) southwest of the cluster center. There is no evidence for the bridge of low--brightness emission detected by Kim et al.~(\cite{kim89}). Its intensity is obviously below the noise level of the 2.675~GHz map.

\begin{figure}[!ht]
\resizebox{\hsize}{!}{\includegraphics[angle=0]{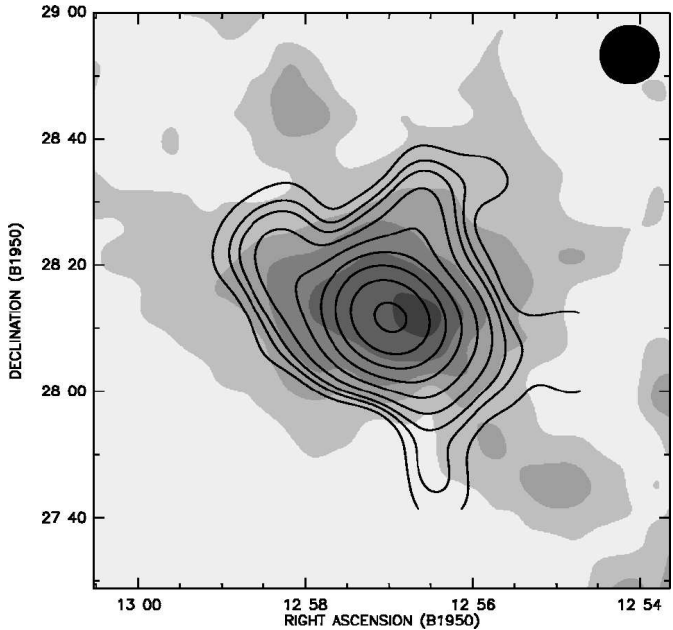}}
\caption{Comparison of the new 2.675~GHz map of the center region of the Coma cluster (contours) with the 1.4~GHz map of Deiss et al.~(\cite{deiss97}) showing the extended diffuse halo emission (greyscale). In both maps the lowest step marks the 1$\sigma$ level. The high--frequency map is smoothed to the 1.4~GHz beamsize ($9\farcm 35$, indicated by the filled circle in the upper right corner) after an approximate (Gaussian) subtraction of the point--like sources. The still visible extensions are remnants of the source removal. Note, we did not remove the two center sources. This overlay shows clearly the smaller extent of the 2.675~GHz emission.
}
\label{fig21-11}
\end{figure}

The central structure of the cluster emission is consistent with observations presented by Wielebinski~(\cite{wieleb78}) and by SST. The map reveals a couple of point sources, mostly with low--level polarized emission. At the center one finds the radiation of the two galaxies NGC~4869 and NGC~4874 convolved to an extended source by the finite telescope beam. Diffuse emission surrounds the central maximum out to about $20\arcmin$. The highest sidelobes of the antenna pattern appear at about 19dB at a distance of about $7\farcm 5$ from the center. Hence, this diffuse structure around the center is not created by sidelobes. Note that we could not apply the CLEAN algorithm (see e.g.~Klein \& Mack \cite{klein95}) because of the lack of a deep antenna pattern at this frequency due to the confusion by background sources. Comparing our map with that of the diffuse halo emission presented by Deiss et al.~(\cite{deiss97}), as done in Figure~\ref{fig21-11}, one clearly sees the smaller extension of the diffuse emission at 2.675~GHz. Our map is large enough to cover the whole halo source and to ensure emission--free area for baseline fitting. Our observations exhibit a large, frequency dependent extension of the halo source Coma~C. Such a behaviour is to be expected in view of the spectral index map of the central region of the Coma cluster of galaxies presented by Giovannini et al.~(\cite{gio93}). Those authors found an inner plateau of about $8\arcmin$ radius with a spectral index $\alpha = 0.8$, whereas outside that central part $\alpha$ strongly steepens and reaches values larger than $1.8$. A comparison of both maps considering the different frequencies and beamsizes and the spectral index shows consistency of the intensity levels of the extended emission in the 2.675~GHz map with the 1.4~GHz map.
%________________________________________________________________

\subsubsection*{The emission at 4.85~GHz}
Figure~\ref{fig6}{\bf b} shows the central part of the Coma cluster of galaxies at 4.85~GHz. Away from the point sources one finds only marginally extended low--brightness emission surrounding the two NGC--galaxies that are resolved at this frequency. In particular, no extended emission above the 3$\sigma$ level is detected.
%________________________________________________________________

\subsection{Integrated diffuse radio flux}

In order to estimate the integrated flux of the halo we first integrated the whole flux surrounding the center down to the zero level using polygons. In case of the 2.675~GHz map we included the bridges of emission towards the east and north of the center, as seen in Fig.~\ref{fig11c}, at the thinnest parts. We subsequently subtracted the fluxes of the point sources lying within the integration area. We calculated the point source flux densities on the basis of the spectral indices provided by Kim (\cite{kim94}). For those sources without any spectral index information we extrapolated the flux densities based on an assumed uniform spectral index of $\alpha=0.8$. All the extrapolated values are below 1$\ldots$2~mJy, so that these faint sources sum up to only about 4\% of the contaminating flux at 2.675~GHz, and about 5\% at 4.85~GHz, respectively. The results are summarized in Table~\ref{tabflux}.
 
\begin{table}[!hb]
\caption{Flux densities of central part of the Coma cluster ($S_{\rm tot}$), of the point sources ($S_{\rm ps}$), and of the diffuse component ($S_{\rm df}$). The errors of $S_{\rm tot}$ depend on the integration area and the uncertainty of the zero level of the maps, those in $S_{\rm ps}$ follow from the calculations; the quadratic combination of both give the error in $S_{\rm df}$. The point--like sources are given by the numbers according to their positions in the list of Kim (\cite{kim94}). For a text file containing these numbers see {\small\tt http://www.mpifr-bonn.mpg.de/div/konti/kim.txt}.}
\label{tabflux}
\begin{center}
\begin{tabular}{ccc|c|l}
\hline
\hline
$\nu$ & $S_{\rm tot}$ & $S_{\rm ps}$ & \boldmath $S_{\rm df}$\unboldmath & \multicolumn{1}{c}{\rm point sources} \\
$[{\rm MHz}]$ & $[{\rm mJy}]$ & $[{\rm mJy}]$ & \boldmath  $[{\rm mJy}]$\unboldmath & \multicolumn{1}{c}{\rm (position in Kim--list)}\\
\hline
&&&& {\rm 135 137 142 145 150 152} \\
&&&& {\rm 158 160 163 166 168 170} \\
&&&& {\rm 171 172 174 183 185 187} \\
\raisebox{1.5ex}[-1.5ex]{$2675$} & \raisebox{1.5ex}[-1.5ex]{$533\pm 21$} & \raisebox{1.5ex}[-1.5ex]{$426\pm 18$} & \raisebox{1.5ex}[-1.5ex]{\boldmath $107\pm 28$\unboldmath} & {\rm191 192 194 196 197 199} \\
&&&& {\rm 200 205 208 210 219 221} \\
&&&& {\rm  223 225} \\
\hline
&&&& {\rm 142 145 150 152 158 160} \\
&&&& {\rm 163 166 168 170 171 172} \\
\raisebox{1.5ex}[-1.5ex]{$4850$} & \raisebox{1.5ex}[-1.5ex]{$256\pm 9$} & \raisebox{1.5ex}[-1.5ex]{$230\pm 8$} & \raisebox{1.5ex}[-1.5ex]{\boldmath $26\pm 12$\unboldmath} & {\rm 185 187 191 192 194 197} \\
&&&& {\rm 200 208 219 221 223} \\
\hline
\end{tabular}
\end{center}
\end{table}

Most of the spectral indices in the Kim list are calculated based on low--frequency ($<$1.6~GHz) measurements. We expect that the spectral index increases with frequency ($S\propto\nu^{-\alpha}$), so that our calculated point source fluxes will be overestimated. Such a behaviour is evident in case of the two central galaxies NGC~4869 and NGC~4874 (numbers 187 and 194 in the Kim list). The catalog contains 4.835~GHz fluxes for both sources given by observations. As shown in Table~\ref{tabngc} their flux densities are much lower than those calculated with the derived spectral indices (resulting from best fits to the data). In assuming a constant spectral index one overestimates the point source flux and subsequently underestimates the flux density of the diffuse component.

\begin{table}
\renewcommand{\arraystretch}{1.2}
\caption{Flux densities of the central galaxies NGC~4869 and NGC~4874. $S_{4.835}^{\rm VLA}$ is the value given by Kim (\cite{kim94}), $S_{4.85}^{\rm calc}$ is that obtained from the spectral index given in the same work, and $S_{4.85}^{\rm Eff}$ is the peak flux of the source in our map. The calculations using the spectral indices, based mainly on low--frequency data, obviously overestimate the fluxes compared to the measurements. The spectral indices will increase with increasing frequency.}
\label{tabngc}
\begin{center}
\begin{tabular}{llll}
\hline
\hline
{\rm N4869:} & $S_{4.835}^{\rm VLA}=97.5\,{\rm mJy}$ & $S_{4.85}^{\rm calc}=119\,\;{\rm mJy}$ & $S_{4.85}^{\rm Eff}=95.0\,{\rm mJy}$ \\
{\rm N4874:} & $S_{4.835}^{\rm VLA}=77.3\,{\rm mJy}$ & $S_{4.85}^{\rm calc}=83.7\,{\rm mJy}$ & $S_{4.85}^{\rm Eff}=72.8\,{\rm mJy}$ \\
\hline
\end{tabular}
\end{center}
\end{table}

%________________________________________________________________

\subsection{The spectrum of Coma~C}
All data available in the literature together with their references are summarized in Table~\ref{tabspec}. In contrast to earlier publications about the flux density of Coma~C one will not find the work by Waldthausen~(\cite{waldthausen1980}). His 5~GHz observation of the region only covers the central part of the cluster and was done to check the morphology of the central sources. In particular Waldthausen did not mention any 5~GHz flux value in his thesis. The often cited value ($<$52~mJy) is Waldthausen's 2.7~GHz upper limit for the Coma~C flux.

The resulting spectrum of the halo source Coma~C is displayed in Fig.~\ref{figspec}. The new measurements of this work are symbolized with filled squares. Despite the fact that our 2.675~GHz point is slightly higher than that of SST, the break in the spectrum, first mentioned by those authors, is visible. The first 4.85~GHz observation supports this spectral steepening. A more detailed discussion of the spectrum will follow in the next chapter.

\begin{table}[!h]
\caption{Integrated flux densities from Coma~C. References: (1) Henning (\cite{henning1989}), (2) Hanisch \& Erickson (\cite{hanisch1980b}), (3) Cordey (\cite{cordey85}), (4) Venturi et al.~(\cite{venturi1990}), (5) Kim et al.~(\cite{kim90}), (6) Hanisch (\cite{hanisch1980a}), (7) Giovannini et al.~(\cite{gio93}), (8) Deiss et al.~(\cite{deiss97}), (9) present paper, (10) Schlickeiser et al.~(\cite{schlickeiser87}).}
\label{tabspec}
\begin{center}
\begin{tabular}{llclc}
\hline
\hline
{\rm frequency} & \multicolumn{3}{c}{\rm flux density} & {\rm references} \\
$[{\rm MHz}]$ & & $[{\rm Jy}]$ && \\
\hline
$30.9$  & $49$   & $\pm$ & $10$    & {\rm 1} \\
$43$    & $51$   & $\pm$ & $13$    & {\rm 2} \\
$73.8$  & $17$   & $\pm$ & $12$    & {\rm 2} \\
$151$   & $7.2$  & $\pm$ & $0.8$   & {\rm 3} \\
$326$   & $3.81$ & $\pm$ & $0.03$  & {\rm 4} \\
$408$   & $2.0$  & $\pm$ & $0.2$   & {\rm 5} \\
$430$   & $2.55$ & $\pm$ & $0.28$  & {\rm 6} \\
$608.5$ & $1.2$  & $\pm$ & $0.3$   & {\rm 7} \\
$1380$  & $0.53$ & $\pm$ & $0.05$  & {\rm 5} \\
$1400$  & $0.64$ & $\pm$ & $0.035$ & {\rm 8} \\
$2675$  & $0.11$ & $\pm$ & $0.03$  & {\rm 9} \\
$2700$  & $0.07$ & $\pm$ & $0.02$  & {\rm 10} \\
$4850$  & $0.03$ & $\pm$ & $0.01$  & {\rm 9} \\
\hline
\end{tabular}
\end{center}
\end{table}

\begin{figure}[!h]
\resizebox{\hsize}{!}{\includegraphics[angle=0]{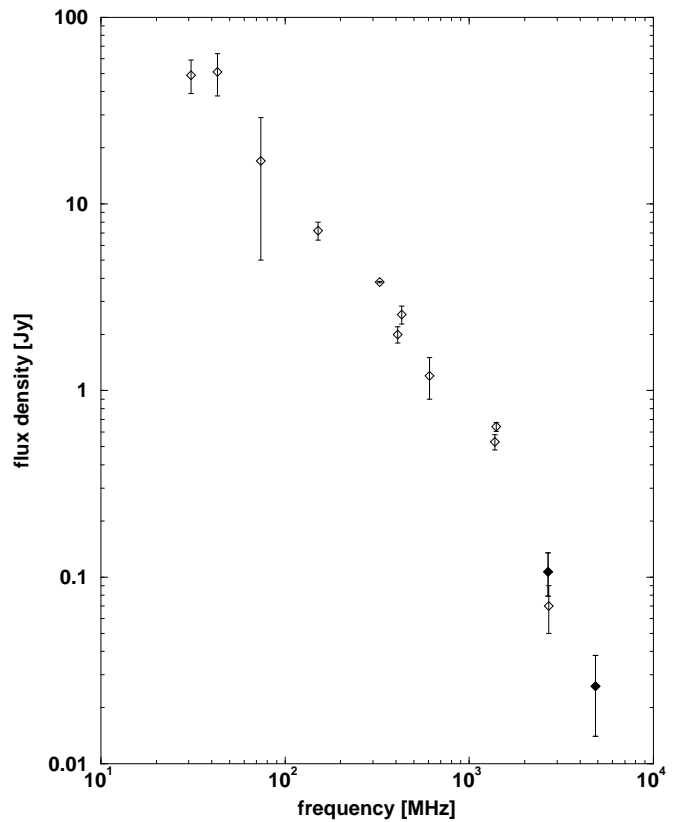}}
\caption{Integrated radio continuum spectrum of the diffuse radio halo source Coma~C. The data and their references are compiled in Table~\ref{tabspec}. The filled dots represent our new observations.}
\label{figspec}
\end{figure}

%________________________________________________________________

\section{Discussion}

\begin{table*}[!t]
\setlength{\tabcolsep}{4pt}
\renewcommand{\arraystretch}{0.8}
\caption{List of the models for cluster halo formation and their best fits to the measurements. Except for the model of Rephaeli~(\cite{reph77}) the free parameters are the best values, with the 90 percent confidence errors. For the model of Rephaeli~(\cite{reph77}) the fit is obviously bad. In particular, the shape of the fit (hence the value of $\chi^{2}$) does not change for $\nu_{\rm K} >$ 100~GHz. The errors displayed in the table are those for the 99 percent confidence.}
\label{tabfits}
\begin{center}
\begin{tabular}{l|l|l|l|l|l}
\hline
\hline
& \multicolumn{3}{l|}{} && \\
Model & \multicolumn{3}{l|}{Primary electron} & Secondary & In--situ \\
& \multicolumn{3}{l|}{} & \raisebox{0.2ex}[-0.2ex]{electron} & \\
\hline
&&&&&\\
Reference & Jaffe~(\cite{jaffe77}) & Rephaeli~(\cite{reph77}) & Rephaeli~(\cite{reph79}) & Dennison~(\cite{denn80}) & Jaffe~(\cite{jaffe77}) \\
&&&&& Roland~(\cite{rol81}) \\
&&&&& Schlickeiser et al.~(\cite{schlickeiser87}) \\
&&&&&\\
\raisebox{2.5ex}[-2.5ex]{Predicted} & \multicolumn{1}{c|}{$\left(\frac{\nu}{\nu_{\rm j}}\right)^{-\frac{\gamma_{0}}{2}}$} & \multicolumn{1}{c|}{$\left(\frac{\nu}{\nu_{\rm j}}\right)^{\frac{1-\gamma_{0}}{2}}\frac{1+\left(\nu_{\rm j}/\nu_{\rm K}\right)^{\frac{1}{2}}}{1+\left(\nu/\nu_{\rm K}\right)^{\frac{1}{2}}}$} & \multicolumn{1}{c|}{$\left(\frac{\nu}{\nu_{\rm j}}\right)^{\frac{2-\gamma_{0}}{2}}\frac{1+\nu_{\rm j}/\nu_{\rm N}}{1+\nu/\nu_{\rm N}}$} & \multicolumn{1}{c|}{$\left(\frac{\nu}{\nu_{\rm j}}\right)^{-\frac{\gamma_{1}}{2}}$} & \multicolumn{1}{c}{$\left(\frac{\nu}{\nu_{\rm j}}\right)^{\frac{3-\Gamma}{2}}exp{{\left(\frac{\nu_{\rm j}^{\frac{1}{2}}-\nu^{\frac{1}{2}}}{\nu_{\rm S}^{\frac{1}{2}}}\right)}}$}\\
\raisebox{5.5ex}[-5.5ex]{spectrum} &&&&&\\
\raisebox{5ex}[-5ex]{$I(\nu)/I(\nu_{\rm j})$} &&&&&\\
$\chi^{2}_{\rm min, red}$ & 1.83 & 1.47 & 0.98 & 1.83 & 0.83 \\
&&&&&\\
Free & $\gamma_{0}=2.69\pm 0.33$ & $\gamma_{0}=3.23^{+0.44}_{-0.48}$ & $\gamma_{0}=4.05^{+0.60}_{-1.12}$ & $\gamma_{1}=2.69\pm 0.33$ & $\Gamma=4.6\pm 0.8$ \\
\raisebox{0.5ex}[-0.5ex]{parameter} && \raisebox{-1.5ex}[1.5ex]{$\nu_{\rm K}=(0.544^{+47.0}_{-0.539})$GHz} & \raisebox{-1.5ex}[1.5ex]{$\nu_{\rm N}=(1.05^{+6.54}_{-1.01})$GHz} && \raisebox{-1.5ex}[1.5ex]{$\nu_{\rm S}=(0.44^{+3.13}_{-0.28})$GHz} \\
&&&&&\\
\hline
\end{tabular}
\end{center}
\end{table*}

The goal of our new observations was the inspection of the spectrum of the halo source. Is the steepening of the spectrum real as claimed by SST? The plot (see Fig.~\ref{figspec}) clearly shows a break towards high frequencies. Both, the SST observation and our measurements yield flux densities below those expected from a power--law extrapolation of the low--frequency range. In order to quantify this behaviour, we followed the same procedure as done by SST. We have fitted the different basic models for halo source formation to the available data points. A $\chi^{2}$--test identifies the model that provides the best fit to the data, and which hence has to be favoured.

In our analysis we considered the primary electron models of Jaffe~(\cite{jaffe77}) and Rephaeli~(\cite{reph77}, \cite{reph79}), the secondary electron model of Dennison~(\cite{denn80}) and the in--situ acceleration model described by Jaffe~(\cite{jaffe77}), Roland~(\cite{rol81}) and SST. Table~\ref{tabfits} gives a summary of the predicted spectra. The notation is identical to that used by SST.

\begin{figure}[!ht]
\resizebox{\hsize}{!}{\includegraphics{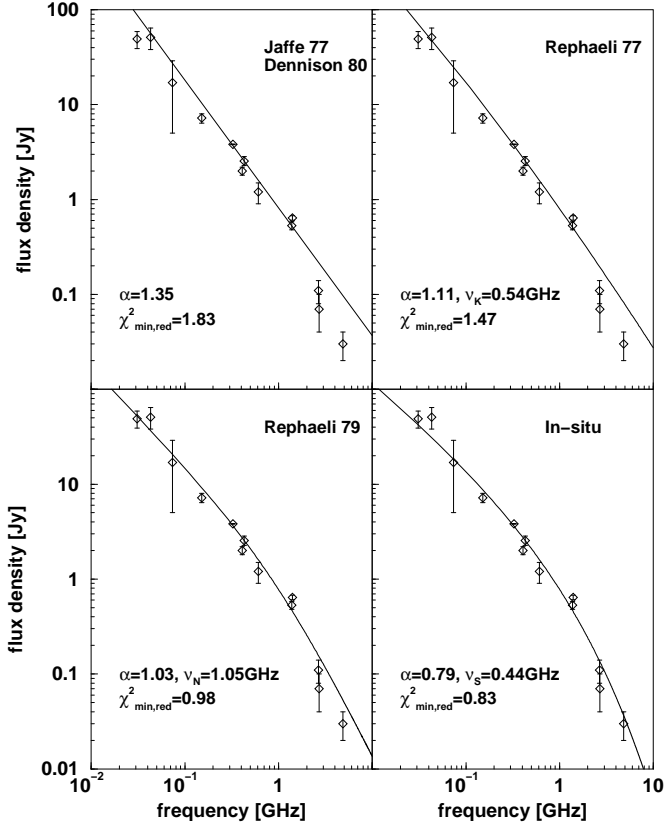}}
\caption{Comparison of the different models for cluster halo formation. The diamonds 
symbolize the data points available in the literature (see Table~\ref{tabspec}). The 
best fits of the models to the data are represented by the solid lines. The resulting 
parameters for the models and the reduced minimized $\chi^{2}$ are displayed in each plot. 
The in--situ acceleration model gives the best fit to the data.}
\label{figplots}
\end{figure}

We normalized the data at $\nu_{\rm j}=0.43$~GHz and determined the most probable values for the parameters using the $\chi^{2}$ technique. Concerning the different measurements with heterogeneous errors we used a weight for the fit of $(1+{\rm d}I/I)$ instead of ${\rm d}I^{2}$ not to overestimate the errors. The last two rows of Table~\ref{tabfits} display the obtained reduced minimized $\chi^{2}$ (=$\chi^{2}_{\rm min}$ per degree of freedom) and the corresponding best values for the free parameters of the models.

Fig.~\ref{figplots} shows the comparison of the calculations. The models of Jaffe~(\cite{jaffe77}) and Dennison~(\cite{denn80}) yield the same predicted spectra and are displayed in the same plot. Our data set contains a couple of new measurements compared to the work of SST. Hence, we obtain different values for the free parameters, but the main conclusion is the same. The in--situ acceleration model gives the best fit to the available data and has to be favoured over the primary and secondary electron models. In particular, the shape of the spectrum reveals a steepening above frequencies of 1~GHz. This was claimed by SST and is proved by the present work.

\subsubsection*{Physical parameters}
The derived shape of the spectrum from the in--situ fit and our map at 2.675~GHz can be used to calculate the magnetic field strength assuming equipartition between the energy density of the relativistic particles and the magnetic field. We model the source as an ellipsoid with axes of the lengths of about 770~kpc, 600~kpc and 600~kpc, yielding an emitting volume of about 1.5$\times 10^{8}$\,kpc$^{3}$. A filling factor of unity is assumed. As the spectral index we use $\alpha=0.79$, which describes the initial power--law in the in--situ fit. In order to avoid any underestimation of the energy of the relativistic particles we have to use a low--frequency flux density given by the fit (we take S$\:\!_{20\rm MHz}$=60~Jy). In order to estimate the total energy we integrate the energy spectrum of the particles from 300~MeV to infinity. This differs from the generally used minimum--energy method (integration over the frequency spectrum from 10~MHz to 10~GHz), but seems more appropriate (see e.g.~Beck et al.~\cite{beck96}). The calculation leads to an equipartition magnetic field of
$$
 B_{\rm eq}=0.57\,(1+k)^{1/(\alpha+3)}\,h_{\rm 75}^{1/(\alpha+3)}\,{\rm \mu G,}\\
$$
where k is the energy ratio of the positively and negatively charged particles and $\alpha$ the spectral index (here 0.79). The resulting total equipartition energy density is
$$
u_{\rm tot}=2.6\times 10^{-14}\,(1+k)^{2/(\alpha+3)}\,h_{\rm 75}^{2/(\alpha+3)}\,{\rm erg/cm^{3}.}
$$
Note that changing k from 1 to 100 changes $B_{\rm eq}$ from 0.68 to 1.9~$\mu$G. These values are in agreement with the equipartition estimate by Giovannini et al.~(\cite{gio93}) as well as with the work of Kim et al.~(\cite{kim90}), who used rotation measures of radio sources projected on the Coma cluster. Nevertheless one has to keep in mind that an estimate of numbers based on the equipartition assumption only provides order of magnitude values. 

With the radio spectrum now well known, we can calculate whether there is a balance between
the internal (relativistic gas) and external (thermal gas) pressure. The latter comes from
X--ray observations. The pressure of the relativistic component is given by
$$
P_{\rm rel} = (\gamma - 1) \times P_{\rm part} + P_{\rm mag},
$$
where $P_{\rm part}$ and $P_{\rm mag}$ are the pressures of particles and fields,
respectively, and $\gamma$ the ratio of specific heats. For relativistic particles,
$\gamma = 4/3$, hence $P_{\rm rel} = 4/3 \times P_{\rm mag}$ in case of energy
equipartition between particles and fields. Inserting numbers, we obtain 
$$
P_{\rm rel}=1.7\times 10^{-14}\,(1+k)^{2/(\alpha+3)}\,h_{\rm 75}^{2/(\alpha+3)}\,{\rm 
dyn/cm^{2}.}
$$
The thermal gas pressure in clusters of galaxies is given by
$$
P(r) = P_{\rm 0}\left(1+\frac{r^2}{r^2_{\rm c}}\right)^{-\frac{3\beta}{2}}
$$
where $P_{\rm 0} = 2n_{\rm 0}kT$ is the central pressure, $r_{\rm c}$ is the core radius, 
$n_{\rm 0}$ is the central electron density, $\beta$ is the ratio of galaxy to gas temperature, 
and $T$ the gas temperature (e.g. Feretti et al.~\cite{feretti92}). At the core radius we thus
obtain a thermal gas pressure 
$$
P_{\rm th} (r = r_{\rm c}) = 4.0\times10^{-11}\,h_{\rm 75}^{1/2}\,\,{\rm dyn/cm^{2}}
$$
using the intracluster plasma properties in Coma as derived by Briel et al.~(\cite{briel92}):
\begin{eqnarray*}
n_{\rm 0} &=& (3.54\pm0.05)\times 10^{-3} h_{\rm 75}^{1/2}\,{\rm cm^{-3}}\\
T &=& 7.8^{+3.8}_{-1.8}\,{\rm keV}\\
r_{\rm c} &=& (10.5\pm 0.6)\,{\rm arcmin}\\
\beta &=& 0.75\pm 0.03.
\end{eqnarray*}
The above calculation shows that the thermal pressure exceeds the relativistic one by a 
factor of about 500, even if we allow for protons and other nuclei in the relativistic gas,
(in which case k = 100). It is thus obvious that the pool of relativistic particles 
forming the radio halo is strongly confined by the ambient thermal gas. This confinement 
was already reported for tailed and low--luminosity radio sources by Morganti et al. ~(\cite{mor88}) and by Feretti et al.~(\cite{feretti92}).

Recently, magnetic field strengths in galaxy clusters have been determined by 
Clarke~et~al.~(\cite{cla01}) using rotation measures of background sources. An average 
intracluster magnetic field strength of 
$B \sim\!5({\ell}/10~{\rm kpc})^{1/2}\,h_{\rm 75}^{1/2}\,{\rm \mu G}$ 
has been obtained by them, using a simple tangled--cell model for the magnetic field 
with a constant coherence length $\ell = 10\,h_{\rm 75}^{-1}\,{\rm kpc}$. Such a high 
value appears to be in conflict with what is obtained if the extreme--ultraviolett (EUV) 
radiation and high energy X--ray (HEX) excess seen in Coma are due to inverse Compton (IC) 
scattering of background photons (see e.g. En{\ss}lin~\&~Biermann~\cite{ens98a}). In the case 
of the Coma halo this would imply magnetic field strengths of a fraction of a $\mu G$, 
much lower than what is implied by the Faraday rotation experiments, and also much lower 
than the values we obtain from the radio synchrotron emission.

The in--situ fit provides both the spectral index of the initial power--law and the cutoff frequency. Using the latter ($\nu_{\rm S}=0.44~$GHz) together with the determined magnetic field, we are able to determine the lifetime of the radiating electrons, $\tau = 5.1\times 10^{7}\,(1+k)^{-3/(2(\alpha+3))}\,h_{\rm 75}^{-3/(2(\alpha+3))}\,{\rm yr}$, taking into account synchrotron and inverse Compton losses. Note that the magnetic field that is equivalent to the cosmic microwave background radiation ($B_{\rm CMB}=4\,(1+z)^{2}\,{\rm \mu G}$) is much larger than the equipartition magnetic field, hence the energy losses via comptonization dominate over the synchrotron losses.

%________________________________________________________________ 
\section{The source 1253+275}\label{sect1253}
\subsection{Introduction}
Our 2.675~GHz observation covers a large region of the Coma cluster including the source 1253+275. This structure was first mentioned by Jaffe~\&~Rudnick~(1979) and subsequently investigated by Andernach~et~al.~(\cite{andernach84}) and Giovannini~et~al.~(\cite{gio85},~\cite{gio91}). 1253+275 is the prototype of a radio relic source. This kind of radio sources can be found at the periphery of galaxy clusters, and they have elongated shapes and often exhibit highly polarized emission. The relevant portion of the 2.675~GHz map that contains the source 1253+275 is enlarged in Fig.~\ref{fig1253}. The extended nature of the source is clearly visible. The map also reveals strong polarized emission within the structure. For the map parameters consult the first row of Table~\ref{tabmaps}.

\begin{figure}[h!]
\resizebox{\hsize}{!}{\includegraphics{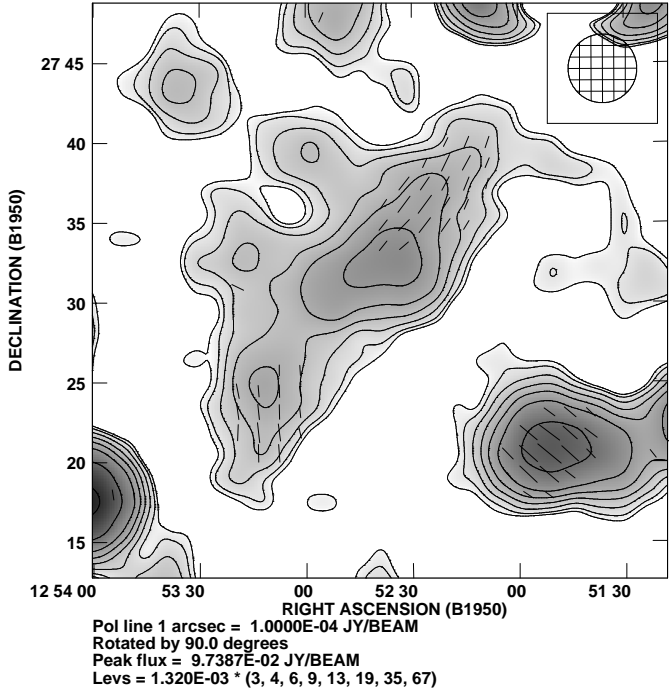}}
\caption{The source 1253+275 at 2.675~GHz. The total power is shown in greyscale and contours, the vectors yield the structure of the magnetic field (E--vectors rotated by 90$\degr$). The lowest contour marks the 3$\sigma$ level. The polarized intensity is represented by the length of the vectors ($1\arcmin = 6$~mJy/beam). The prominent source in the south--west corner is NGC~4789.}
\label{fig1253}
\end{figure}

Since the source 1253+275 appears to lack any optical counterpart (Andernach~et~al.~\cite{andernach84}), its nature is still a puzzle. Giovannini et al.~(\cite{gio85}) initially deemed the cluster member IC3900 to be the source of the plasma, but later on (Giovannini et al.~\cite{gio91}), led by the detection of a bridge of low brightness emission (Kim~et~al.~\cite{kim89}), they interpreted 1253+275 as a part of the large--scale radio emission present in the cluster--wide region of the Coma cluster and Coma--A1367 supercluster. En{\ss}lin et al.~(\cite{ens98b}) favoured NGC~4789 as the possible production site of the relativistic particles. Particles released from this galaxy should feel the large--scale matter flow, directed towards the Coma cluster center. At the contact surface with the dense intracluster medium a shock should be present, thus forcing the particles to emit detectable synchrotron radiation and forming the extended radio source.
 
%________________________________________________________________
\subsection{Morphology}
Our map of 1253+275 is in agreement with that of Andernach~et~al.~(\cite{andernach84}), at the same frequency. The source is extended for $\sim$800~kpc along the major axis (position angle $\sim 135\degr$), perpendicular to the direction towards the cluster center. The maximum extent along the minor axis is $\sim$450~kpc. Linear polarization is seen in the northern and southern part. Andernach~et~al.~(\cite{andernach84}) obtained internal and foreground rotation measures of zero, so that rotation of the polarization vectors by 90$\degr$ discloses the projected orientation and structure of the magnetic field. It can be seen in Fig.~\ref{fig1253} that the magnetic field is aligned parallel to the major axis of 1253+275, which fits well into the picture of a shock in this region. Such a shock should be directed towards the cluster center. Upstream to the shock, the plasma is compressed and the magnetic field is aligned perpendicular to the shock direction. The sharp edge of the structure towards the south--west is an additional hint at a compression of the present plasma.

%________________________________________________________________
\subsection{Integrated diffuse radio flux}
The radio flux of the diffuse emission of source 1253+275 was obtained by integrating all emission within the 3$\sigma$ level. The contribution of point sources was calculated using the source list and the spectral index values provided by Giovannini~et~al.~(\cite{gio91}). The resulting total flux of the extended structure at 2.675~GHz (point sources subtracted) is $112\pm 10$~mJy. Table~\ref{tabpol} contains the relevant data, where the contributions have been calculated for the northern and southern part of the source (corresponding to regions of polarized emission at the 3$\sigma$ level). The polarization properties are also listed.
 
\begin{table}[t!]
\caption{Compilation of radio data for 1253+275. $S_{\rm tot}$ is the total flux density of the diffuse component, $S_{\rm pol}$ its polarized flux, $P$ is the degree of polarization, and $\psi$ the polarization angle.}
\label{tabpol}
\begin{center}
\begin{tabular}{lllllllll}
\hline
\hline
\multicolumn{1}{c}{region} && \multicolumn{1}{c}{$S_{\rm tot}$} && \multicolumn{1}{c}{$S_{\rm pol}$} && \multicolumn{1}{c}{$P$} && \multicolumn{1}{c}{$\psi$} \\
&& \multicolumn{1}{c}{$[{\rm mJy}]$} && \multicolumn{1}{c}{$[{\rm mJy}]$} && \multicolumn{1}{c}{$[\%]$} && \\
\hline
1253+275 entire && $112\pm 10$ && $29\pm 4$ && $26\pm 4$ && \\
1253+275 north && $\;\:33\pm \;\:2$ && $10\pm 1$ && $30\pm 4$ && $\;\:56\degr$ \\
1253+275 south && $\;\:20\pm \;\:2$ && $\;\:8\pm 1$ && $40\pm 6$ && $\;\:90\degr$ \\
NGC~4789 && $\;\:71\pm \;\:7$ && $11\pm 1$ && $15\pm 2$ && $140\degr$ \\
\hline
\end{tabular}
\end{center}
\end{table}

\begin{figure}[!t]
\resizebox{\hsize}{!}{\includegraphics{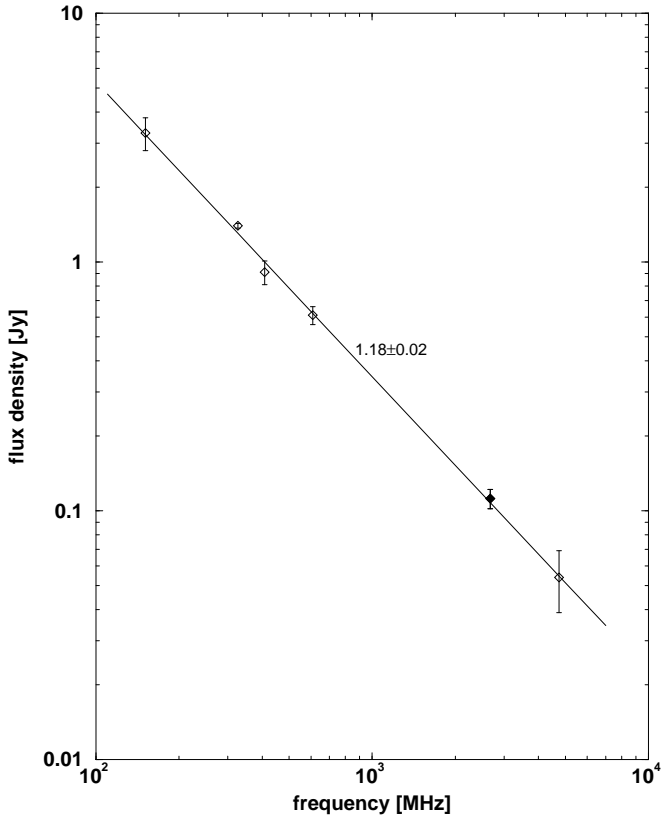}}
\caption{Integrated radio continuum spectrum the extended source 1253+275. The line symbolizes the best power--law fit to the data, with a spectral index of $\alpha = 1.18\pm 0.02$.}
\label{fig1253spec}
\end{figure}

\begin{table}[h!]
\caption{Integrated flux densities of source 1253+275.}
\label{tab1253spec}
\begin{center}
\begin{tabular}{llclll}
\hline
\hline
{frequency} & \multicolumn{3}{c}{fluxdensity} && {reference} \\
$[{\rm MHz}]$ & & $[{\rm Jy}]$ &&& \\
\hline
$151$   & $3.3$   & $\pm$ & $0.5$   && Cordey (\cite{cordey85}) \\
$326$   & $1.40$  & $\pm$ & $0.03$  &&  Giovannini~et~al.~(\cite{gio91}) \\
$408$   & $0.91$  & $\pm$ & $0.10$  && Ballarati (\cite{ballarati81}) \\
$610$   & $0.611$ & $\pm$ & $0.05$  && Giovannini~et~al.~(\cite{gio91}) \\
$2675$  & $0.112$ & $\pm$ & $0.010$ && {present paper} \\
$4750$  & $0.054$ & $\pm$ & $0.015$ && Andernach et al.~(\cite{andernach84}) \\
\hline
\end{tabular}
\end{center}
\end{table}

The flux density of 1253+275 obtained by us is lower than that of Andernach~et~al.~(\cite{andernach84}) if one subtracts the same point source contribution from their number. Our new measurement fits well to the other data available in the literature. The spectrum of the source is plotted in Fig.~\ref{fig1253spec}, with the data listed in Table~\ref{tab1253spec}. The best fit to the data yields a spectral index of $\alpha=1.18\pm0.02$. This is the same result as that of Giovannini~et~al.~(\cite{gio91}), though with a smaller error.

Using the equipartition assumption we derive a magnetic field strength of $B_{\rm eq}=0.56\,(1+k)^{1/(\alpha+3)}\,h_{\rm 75}^{1/(\alpha+3)}\,{\rm \mu G}$ in 1253+275, which implies a total energy density of $u_{\rm tot}=2.5\times 10^{-14}\,(1+k)^{2/(\alpha+3)}\,h_{\rm 75}^{2/(\alpha+3)}\,{\rm erg/cm^{3}}$. Here we use a filling factor of 1, the spectral index of $\alpha$=1.18 and the flux density of  our measurement at 2.675~GHz. The emitting volume of about 1.4$\times 10^{8}$\,kpc$^{3}$ results from treating the source as a flat cylinder of height 650$''$ (280~kpc, orientated towards the cluster center) and diameter 1850$''$ (800~kpc, parallel to the major axis). Our numbers differ only slightly from those used by Andernach~(\cite{andernach84}) and Giovannini et al.~(\cite{gio91}).
 
%________________________________________________________________
\section{Summary}
We presented new single--dish radio observations of the Coma cluster of galaxies using the Effelsberg 100--m telescope. Our investigations were focused on the diffuse extended radio sources in the cluster. The radio halo source Coma~C was observed at two frequencies (2.675 and 4.85 GHz). After subtracting the contribution of point sources we obtained new determinations of the flux densities of the diffuse structures. These verify the strong steepening of the radio continuum spectrum of Coma~C at frequencies above 1~GHz, as first noticed by Schlickeiser et al.~(\cite{schlickeiser87}). In comparing the three basic models for radio halo formation (primary electron model, secondary electron model and in--situ acceleration model), similar to the work done in Schlickeiser et al., we find that the latter most probably reflects the process at work in the Coma cluster. Fits of the models to the available data using a $\chi^{2}$ test yield the best result for the in--situ acceleration model. This describes the spectral steepening also quantitatively.

Our 2.675~GHz map is extended enough to cover the entire diffuse emission. Our map is in good agreement with that published by Schlickeiser et al., thus settling a long--standing uncertainty (see the discussion of Deiss et al.~\cite{deiss97}). Our 4.85~GHz measurement corroborates the spectral steepening towards high frequencies. Part of the previous confusion results from a misquotation in a couple of papers concerning the flux density of Coma~C at 5~GHz. The often cited Ph.D.~thesis of Waldthausen~(\cite{waldthausen1980}) does not provide any 5~GHz flux density of Coma~C. The value of less than 52~mJy, used in the citations as the total flux density at 5~GHz, is Waldthausen's 2.7~GHz upper limit to the flux density of the extended component. Our 4.85~GHz observation provides the first firm measurement of the flux density of Coma~C at this frequency.

Based on the equipartition assumption we calculate a field strength $B_{\rm eq}=0.57\,(1+k)^{1/(\alpha+3)}\,h_{\rm 75}^{1/(\alpha+3)}\,{\rm \mu G,}$ and a total energy density of $u_{\rm tot}=2.6\times 10^{-14}\,(1+k)^{2/(\alpha+3)}\,h_{\rm 75}^{2/(\alpha+3)}\,{\rm erg/cm^{3}}$ for in Coma~C.

Our 2.675~GHz map includes the region in which Kim et al.~(\cite{kim89}) detected a bridge of low--brightness emission. Our map does not show any significant feature to be identified with the bridge. Its brightness is probably too low at 2.675~GHz to be detected.

We derive a new 2.675~GHz flux density for the source 1253+275, an extended diffuse radio relic located at the south--western periphery of the Coma Cluster. Using this new measurement, we calculate a spectral index of $\alpha$=1.18$\pm 0.02$, a value that is more accurate than previous determinations. For 1253+275 we estimate an equipartition magnetic field strength $B_{\rm eq}=0.56\,(1+k)^{1/(\alpha+3)}\,h_{\rm 75}^{1/(\alpha+3)}\,{\rm \mu G}$ and a total energy density of $u_{\rm tot}=2.5\times 10^{-14}\,(1+k)^{2/(\alpha+3)}\,h_{\rm 75}^{2/(\alpha+3)}\,{\rm erg/cm^{3}}$.

%________________________________________________________________

\begin{acknowledgements}
MTH thanks Dr.~Ch.~Nieten for his assistance in programming the fit procedure. Dr.~R.~Beck is acknowledged for critical reading of the manuscript.
\end{acknowledgements}

%________________________________________________________________

%________________________________________________________________

\end{document}